Enhancing carrier transfer properties of Na-rich anti-perovskites, $Na_4OM_2$ with tetrahedral anion groups: an evaluation through first-principles computational analysis.

Shenglin Xu [a,b], Qinfu Zhao [a], Ronglan Zhang [b], Bingbing Suo [a], Qi Song [a,*]

[a] Shaanxi Key Laboratory for Theoretical Physic Frontiers, Institute of Modern Physics, Northwest University, Xi'an, 710069, PR China

[b] Key Laboratory of Synthetic and Natural Functional Molecule Chemistry of Ministry of Education, College of Chemistry and Materials Science, National Demonstration Center for Experimental Chemistry Education (Northwest University), Northwest University, Xi'an, Shaanxi, 710069, PR China

*E-mail: songq@nwu.edu.cn

**ABSTRACT**

The practical application of Na-based solid-state electrolytes (SSEs) is limited by their low level of conduction. To evaluate the impact of tetrahedral anion groups on carrier migration, we designed a set of anti-perovskite SSEs theoretically based on the previously reported $Na_4OBr_2$, including $Na_4O(BH_4)_2$, $Na_4O(BF_4)_2$, and $Na_4O(AlH_4)_2$. It is essential to note that the excessive radius of anionic groups inevitably leads to lattice distortion, resulting in asymmetric migration paths and a limited improvement in carrier migration rate. $Na_4O(AlH_4)_2$ provides a clear example of where $Na^+$ migrates in two distinct environments. In addition, due to different spatial charge distributions, the interaction strength between anionic groups and $Na^+$ is different. Strong interactions can cause carriers to appear on a swing, leading to a decrease in conductivity. The low conductivity of $Na_4O(BF_4)_2$ is a typical example. This study demonstrates that $Na_4O(BH_4)_2$ exhibits remarkable mechanical and dynamic stability and shows ionic conductivity of $1.09 \times 10^{-4}$ S cm$^{-1}$, two orders of magnitude higher than that of $Na_4OBr_2$. This is attributed to the expansion of the carrier migration channels by the anion groups, the moderate interaction between carriers and anionic groups, and the "paddle-wheel" effect generated by the anion groups, indicating that the "paddle-wheel" effect is still effective in low-dimensional anti-perovskite structures, in which atoms are arranged asymmetrically.

## Introduction

Over the past three decades, the demand for electrochemical energy storage has accelerated the development of solid-state electrolytes (SSEs). Although significant advancements have been made in lithium-ion batteries (LIBs), including enhanced energy density and safety, the high cost of lithium remains a significant hurdle in reducing LIBs' cost. Consequently, sodium-ion batteries (SIBs) have become a promising alternative owing to their abundance and low cost. Nonetheless, the SIBs suffer from several constraints such as lower energy density, shorter lifespan, higher internal resistance, and lower ionic conductivity, which severely limit the output power of these batteries. As such, it is paramount to search for and design SIBs materials with high ionic conductivity and excellent chemical/electrochemical stability.

Recent reports have highlighted inorganic sodium superionic conductors, which have promoted the exploration of anti-perovskite $Li_3OCl$ sodium analogs. [1] These studies have shown that anti-perovskite $Na_3OX$ structures exhibit strong sodium ionic conductivity at high temperatures. [2] As two-dimensional materials have exceptional physicochemical properties, recent studies have focused on exploring the rapid ion conduction mechanism of low-dimensional anti-perovskites through theoretical calculations. These investigations have divulged that the phonon softening effect in the low-dimensional structure can cause the rapid migration of cations and increase ionic conductivity (about $10^{-4}$ S $cm^{-1}$). Further experimentation indicates that the primary migration channel of $Na^+$ in multi-layered $Na_4OI_2$ occurs between adjacent $Na^+$ cations residing in the $NaO_6$ octahedron. [3] The polarized $I^-$ anions join the $Na^+$ hopping pathway between the layered anti-perovskites' separated octahedra along the c-axis. [4] These revelations suggest that subsequent structural modifications can enhance $Na^+$ migration channels and heighten ionic conductivity. It is crucial to note that weak interactions between polarizable anions also impact ionic conductivity.

Research has revealed that by replacing monovalent anions with anionic groups in $Na_3OX$ (X=Cl, Br, and I), it is feasible to elevate the ionic conductivity by an astonishing four orders of magnitude.[5] Cubic anti-perovskite $Na_3OBH_4$ possesses an ionic conductivity of $4.4 \times 10^{-3}$ S $cm^{-1}$ and $1.1 \times 10^{-2}$ S $cm^{-1}$ at ambient temperature and 328 K, respectively, representing an impressive increase.[6] A mixed phase of $Li_3S(BF_4)_{0.5}Cl_{0.5}$ and the $Li_3SBF_4$ proposal exhibits a room temperature conductivity of approximately 0.1 S $cm^{-1}$, which is comparable to organic electrolytes. [7] The remarkable improvement in conductivity is attributed to the cubic anti-perovskite crystal structure, as well as the rotational nature of its anion groups.

In comparison with $Na_3OX$ and other three-dimensional anti-perovskite structures, $Na_4OBr_2$ exhibits a distinctive low-dimensional, multi-layer anti-perovskite structure and enjoys a greater carrier concentration, making it a promising candidate to adjust the structure in enhancing ionic conductivity.[8] To substitute $Br^-$, anionic groups such as $[BH_4]^-$, $[BF_4]^-$, and $[AlH_4]^-$ were used due to their variations in size, interaction, and chemical properties.

**Methods**

The Vienna ab initio simulation package (VASP) was utilized to conduct first-principles calculations, employing Perdew-Burke-Ernzerhof (PBE) exchange and correlation functionals and projector-augmented-wave (PAW) pseudopotentials.[9−15] A 10×10×10 k-mesh grid and a cutoff energy of 600 eV have been utilized. [16] The electronic self-consistent field and the forces acting on atoms have a convergence threshold of $1.0\times10^{-7}$ eV and $1.0\times10^{-4}$ eV $Å^{-1}$, respectively. All geometries were fully relaxed until the residual Hellmann−Feynman forces acting upon each atom were less than 0.01 eV $atom^{-1}$, ensuring consistent and stable results. In addition, the van der Waals interactions were corrected using the DFT-D3 method proposed by Grimme et al.[17] In analyzing the crystal orbital Hamiltonian population (COHP), an integrated crystal orbital Hamiltonian population (ICOHP) acts as an ab initio measure of the covalent bond strength.[18]

A 2×2×1 supercell with a 3×3×2 k-point grid to compute the phonon dispersion and 2nd interatomic force constant (2nd IFC) in conjunction with the VASP and PHONOPY packages. [19−22] One vacancy is created for every 32 sodium atoms, and the sodium: sodium vacancy ratio in the lattice is 1:0.03. For $Na_4OBr_2$, $Na_4O(BH_4)_2$, $Na_4O(BF_4)_2$, and $Na_4O(AlH_4)_2$, the formation energy of a sodium vacancy is approximately 0.941 eV, 0.348 eV, 0.586 eV, and 0.886 eV in $Na_4OBr_2$, $Na_4O(BH_4)_2$, $Na_4O(BF_4)_2$, and $Na_4O(AlH_4)_2$, respectively. These values are much lower than the sodium interstitial value in $Na_4OBr_2$, $Na_4O(BH_4)_2$, $Na_4O(BF_4)_2$, and $Na_4O(AlH_4)_2$, with values of 1.294 eV, 1.711 eV, 1.805 eV and 1.996 eV respectively. To determine migration pathways, we used the climbing image nudged elastic band method (CI-NEB) in conjunction with the dimer method.[23,24] Supercells have been utilized to minimize carrier interaction with their replicas during migration.

**Results and discussions**

*Structural Properties*

Drawing on the previously reported crystal structure of $Na_4OBr_2$, we have proposed theoretical structures for two-dimensional anti-perovskites $Na_4O(BH_4)_2$, $Na_4O(BF_4)_2$, and $Na_4O(AlH_4)_2$ by substituting $[BH_4]^-$, $[BF_4]^-$, and $[AlH_4]^-$ anion groups in place of the halogen $Br^-$, as depicted in Figure 1. In order to alleviate the extensive internal stress imposed by anion substitution, a sequence of lattice parameter optimization and ionic position relaxation steps must be executed. The lattice parameters and a variety of bond distances have been optimized by using the PBE-D3 method and the results are presented in Table 1.

The optimized structural parameters of $Na_4OBr_2$ exhibit only 4% and 1% errors for the *c*-axis and $a$-axis, respectively, compared to experimental values, bearing testimony to the credibility of both the method and its accompanying parameters.[25] Due to the anisotropic and rotatable nature of the substituted anionic groups, the symmetry of the space group in these theoretical structures is diminished, as illustrated in Figure 1. As the radius of the anionic group increases, the lattice parameters along the *a* and *b* axes are no longer equivalent, albeit they remain vertical and display anisotropy on the *ab* plane. The average distance between $Na^+$ and the center of the anionic group expands, suggesting that the transport channels for carriers must inevitably expand too. Notably, while $Na_6O$ tilts in Figure 1, the coordination number remains unaltered, thereby hinting at the preservation of the physicochemical properties of these novel structures.

The Goldschmidt tolerance factor ($t$) has been extensively employed to anticipate the formability of prospective anti-perovskite structures, which is defined as:[26]

$$t = \frac{R_{Na}+R_M}{\sqrt{2}(R_O+R_M)} \quad (1)$$

where $R_{Na}$ is the ionic radius of sodium cations, Ro is the radius of oxygen, and $R_M$ is the radii of anionic groups, respectively, as per the Shannon and Prewitt (SP) method.[27] [28] The radii significantly differ by the balancing ion that is employed for the anionic groups. An effective radius for the group has been suggested by Kieslich et al. predicated on a collection of effective ionic radii, wherein the radius of a group ($XY_n$) is defined as the summation of the X–Y bond length and the radius of Y.[29,30] These semi-empirical radii may be utilized to gauge tolerance factors and ascertain the stability of new structures.

The calculated Goldschmidt tolerance factors are presented in Table 2. It suggests that the ideal tolerance factor for cubic perovskite structures should range between 0.8 and 1.0, which is also applicable to anti-perovskite structures. A tolerance factor exceeding 1.0 leads to the formation of a twisted anti-perovskite structure with slanted octahedra. Since the ionic radius of the $[BH_4]^-$ group closely resembles that of Br-, substitutions with $[BH_4]^-$ are unlikely to affect the structure. The larger radii of $[BF_4]^-$ and $[AlH_4]^-$ increase *t* slightly over to 1, altering the crystal structure from a tetragonal lattice to an orthogonal system. The octahedral factor μ is introduced as a supplement to the Goldschmidt tolerance factor,[31] which is defined as:

$$\mu = \frac{R_O}{R_M} \qquad (2)$$

The calculated results are listed in Table 3. All values of μ are greater than 0.414, indicating that the stability of $OM_6$ has not qualitatively changed due to the introduction of anion clusters. Therefore, all of these candidate structures are stable.[32]

To further evaluate structural stability and compare the mechanical properties of each candidate, the elastic constants were calculated according to the generalized Hooke's law and are enumerated in Table 3.[33] All eigenvalues of $C_{ij}$ were positive, satisfying the critical Cauchy-Born criterion for mechanical stability of the corresponding symmetry. Table 3 also shows the calculated mechanical properties of polycrystalline materials, including bulk modulus and shear modulus. The presence of substituted anionic groups of $[BH_4]^-$ and $[BF_4]^-$ improved their respective mechanical properties. However, the introduction of $[AlH_4]^-$ caused significant structural distortion, resulting in $Na_4O(AlH_4)_2$ being relatively soft.

Various high-symmetry directions in the Brillouin zone, as well as the total and projected phonon density of states have been plotted in Figure 2. It can be observed that each candidate exhibits dynamic stability, given that there are no imaginary frequencies. Previous reports on $Li_3OCl$ and $Na_3OCl$ have shown that three-dimensional perovskite structures are thermodynamically and dynamically unstable at low temperatures, and can only exist in metastable state at high temperatures, high pressures or distorted structures. Our results suggest that lower dimensions are more conducive to anti-perovskite structure stability. [34] Two pronounced phonon gaps are discernible in the phonon dispersion spectra of $Na_4OM_2$, partitioning the whole phonon bands into three frequency regions. The presence of anionic groups enriches the atomic vibration modes, which is similar to the spectral range of the three-dimensional perovskite materials $Na_3O(BF_4)$ and $Na_3O(AlH_4)$. [30] Additionally, the contributions of Na and O atoms exist in the low-frequency region (0.0~15.0 THz), and are almost unaffected by changes in M anions. Introduction of $[BH_4]^-$ and $[AlH_4]^-$ groups, the H atoms in the outer layer of the anion group contribute to the peak of the high-frequency region, which is indicative of the obvious optical branches, as illustrated in Figure 2 (b)(d). However, in $[BF_4]^-$, the high-energy vibration is attributed to the B atom located in the center of the anionic group, corresponding to the frequency region (30.0~40.0THz).

*Electronic structures*

A solid electrolyte material that is suitable for use in batteries should not only exhibit high ionic conductivity but should also maintain electronic insulation. Electronic conductivity is a critical parameter for SSEs, and it is determined by the band structure, especially the energy gap between the valence and conduction bands. Band structures, corresponding density of states (DOS) and partial DOS for each element have been calculated and displayed in Figure 3. As shown in Figure 3, the valance band maximum (VBM) of these structures is mainly contributed by the 2*p* orbitals of O, and the factors affecting the band gap are the energy and contribution of the atomic orbitals corresponding to the conduction band minimum (CBM). The CBM of $Na_4O(BH_4)_2$ is narrower than that of the other models, and the PDOS indicates that the corresponding eigenstate of this band is mainly composed of the s orbitals of $Na^+$ cations localized on the lattice points, as shown in Figure 3(b). This band has strong electron localization and a relatively large effective mass, so the energy of the CBM is higher, resulting in a larger band gap and the worst electron conductivity. The band gap of $Na_4OBr_2$ measures 2.128 eV, but this value increases to 3.382 eV, 2.850 eV, and 3.100 eV when $[BH_4]^-$, $[BF_4]^-$, and $[AlH_4]^-$ anions substituting the $Br^-$, respectively. Thus, it is clear that substituting halogen with anionic groups leads to a decrease in electron conductivity.

To verify the chemical bond strength predictions, the COHP sum of states below the Fermi level can be used, as depicted in Figure 4. According to the bonding states' area in Figure 4, the B-F and Al-H bonds within the anionic group appear relatively weaker than the B-H bonds. It is important to determine the amount of charge transfer in a scientific and quantitative manner. Using Bader charge analysis from Table 4, it is clear that different anionic functional groups yield a similar impact on the charge distribution of ions in the crystal.[35]

*Ionic conductivity*

This section discusses the $Na^+$ migration in $Na_4OM_2$. The activation energy of $Na^+$ diffusion and its pathways are crucial when assessing $Na_4OM_2$ as a plausible material for sodium-ion batteries. Both ab initio Molecular dynamics (AIMD) and NEB methods have been used to calculate $Na^+$ conductivity based on density functional theory. [36–38] AIMD simulations mainly focus on the dynamic behavior and thermal properties of carriers. While NEB method finds the lowest energy path during the migration, to obtain the migration energy of carriers and calculate the conductivity of the carriers. This method focuses on the interaction between carriers and their coordination environment. In this work, CI-NEB and Dimer methods are utilized to compute the $Na^+$ migration barrier in each model structure, as depicted in Figure 5. For all CI-NEB calculations, three images were inserted between the initial and final states. Subsequently, the Dimer method is implemented to determine the migration states. The corresponding diffusion coefficient and ionic conductivity were determined utilizing the Arrhenius Equation, and these results are set forth in Table 5.

$$D_s = D_0 e^{-\frac{E_a}{k_b T}} = g a^2 v e^{-\frac{E_a}{k_b T}} \tag{3}$$

$$\sigma = \frac{n e^2 Z^2}{k_b T} D_s \tag{4}$$

where $g$ represents the geometrical attribute, $a$ denotes the migration distance, $v$ signifies the attempt frequency, $E_a$ embodies the sodium defects' migration gate, $n$ is the defect concentration, equal to 1/32. $Z$ characterizes the carrier charge, $e$ represents the elementary charge, $k_b$ is the Boltzmann constant, and $T$ correlates to the thermodynamic temperature fixed at 298K.

The $Na^+$ migration barrier in $Na_4OBr_2$ measures 0.43 eV, whereas the integration of $[BH_4]^-$ reduces this barrier value to 0.32 eV. Conversely, in $Na_4OM_2$, the $Na_6O$ octahedra shape a migration layer as $Na^+$ cations, connected at equatorial locations to other octahedra, permit migration. In the course of migration, $Na^+$ must follow a bent path, avoiding steeper potential energy curves, by transiting through migration gates (MGs) generated by $O^{2-}$ and two anions, as displayed in Figure 6. [39] The MGs in $Na_4OBr_2$ comprises three rigid, spherical anions, whereas $Na_4O(BH_4)_2$'s MGs includes one $O^{2-}$ anion and two that rotate and vibrate $[BH_4]^-$. The anionic groups diminish the repulsion that exists between carriers and MGs, due to the mobile H atoms, which reduce the migration barriers adeptly. During carrier migration, the rotation of the $[BH_4]^-$ tetrahedra, aligning the surface of the tetrahedra towards the migrating cation as much as possible, acts as the "paddlewheel effect".[30,40,41]

Figure 5 demonstrates that the insertion of anionic groups $[BF_4]^-$ and $[AlH_4]^-$ leads to deviations in the initially identical migration channels. Figure 7 provides further clarification on the coordination surroundings of charge carriers during migration. As depicted in Figures 7(a)(b), it is evident that the coordination environments for the two selected paths, PATH_A and PATH_B, are equivalent during $Na^+$ migration in the *ab* plane of both $Na_4OBr_2$ and $Na_4O(BH_4)_2$. With the introduction of larger-sized $[BF_4]^-$ and $[AlH_4]^-$ anionic groups, the symmetry disruption progresses from *Pnnm* to *Fmm2*, resulting in $Na^+$ experience unique coordination surroundings across diverse migration channels, as illustrated in Figure 7(c)(d).

The disruption of system symmetry caused by large-sized anionic clusters is reflected in changes in the distances between atoms and the strength of their interaction forces. The ICOHP values in Table 6 elucidate the interaction forces between sodium and the outermost atoms of the anionic clusters. Higher ICOHP values for sodium with the outer $F^-$ atoms of the $[BF_4]^-$ anionic cluster and the outer $H^-$ atoms of the $[AlH_4]^-$ anionic cluster (PATH_A) indicate a stronger attraction between these atoms and sodium, which is consistent with the coordination environment of sodium shown in Figures 7(c) and 7(d). Excessively strong interaction forces hinder sodium migration, resulting in low ionic conductivity, as listed in Table 5. However, the disruption of symmetry does not increase the interaction forces between sodium and H in the PATH_B pathway of $Na_4O(AlH_4)_2$, as shown in Figure 7(d). The substitution of large size $[AlH_4]^-$ will not meet the limitations of Goldschmidt tolerance factor and octahedral factor, thus inevitably reducing the symmetry of the structure, and hinder sodium migration. The symmetry disruption leads to irregular atomic arrangements, making it difficult to control the changes in carrier coordination environments within the same system. While the coordination environment of sodium in the PATH_B pathway is similar to that in $Na_4O(BH_4)_2$. This is attributed to the appropriate size of the $BH_4^-$ anion groups. An excessively large volume would instead hinder sodium migration. Additionally, the outer atoms of the introduced ionic clusters should not have excessively strong interactions with sodium.

**Conclusions**

In this work, to assess the impact of tetrahedral anionic groups on $Na^+$ migration, we designed a set of complex structures of $Na_4O(BH_4)_2$, $Na_4O(BF_4)_2$, and $Na_4O(AlH_4)_2$ founded on the reported $Na_4OBr_2$. Elastic constants and phonon calculations confirmed the mechanical and thermodynamic stability of each structure. CI-NEB and DIMER methods were utilized to calculate the migration processes of carriers in each candidate, and the ionic conductivity of carriers was determined by combining migration distances.

Substituting $Br^-$ with $[BH_4]^-$ anionic groups, $Na_4O(BH_4)_2$ has significantly increased the ionic conductivity of $Na^+$ cations by about two orders of magnitude. This is largely attributed to the effective paddle-wheel mechanism of $[BH_4]^-$ anionic groups, which reduces migration energy barriers by diminishing the repulsion between carriers and MGs during migration. However, the excessive ion volume of $[BF_4]^-$ and $[AlH_4]^-$ anionic groups, combined with robust interactions, causes cell distortion when vacancies are present in the system, resulting in significant differences in migration paths within the same plane and unstable conductivity. Our results demonstrate that $Na_4O(BH_4)_2$, in the *Pnnm* space group is both mechanically and thermodynamically stable, the ionic conductivity is two orders of magnitude higher than that of $Na_4OBr_2$, indicating that $Na_4O(BH_4)_2$ could be a superior solid-state electrolyte. Additionally, theoretical analyses reveal that excessive interaction and volume can result in negative effects, culminating in decreased ion conductivity. High-throughput calculations and machine learning will enable us to sift through various anionic groups and strive to discover optimal solutions in following researches.

**Acknowledgments**

This work was supported by the Natural Science Foundation of Shaanxi Province of China (No. 2020JQ-568), the Natural Science Basic Research Plan in Shaanxi Province of China (No. S2020-JC-QN-0623), the National Natural Science Foundation of China (No.

2022GY-375), the Key Laboratory Research and Establish Program of Shaanxi Education Section (No. 17JS130) and the Double First-class University Construction Project of Northwest University.

**Table of Contents entry**

Sodium ions migrate through migration gates (MGs) in $Na_4O(BH_4)_2$ and cause rotation of the $[BH_4]^-$ clusters that make up MGs.

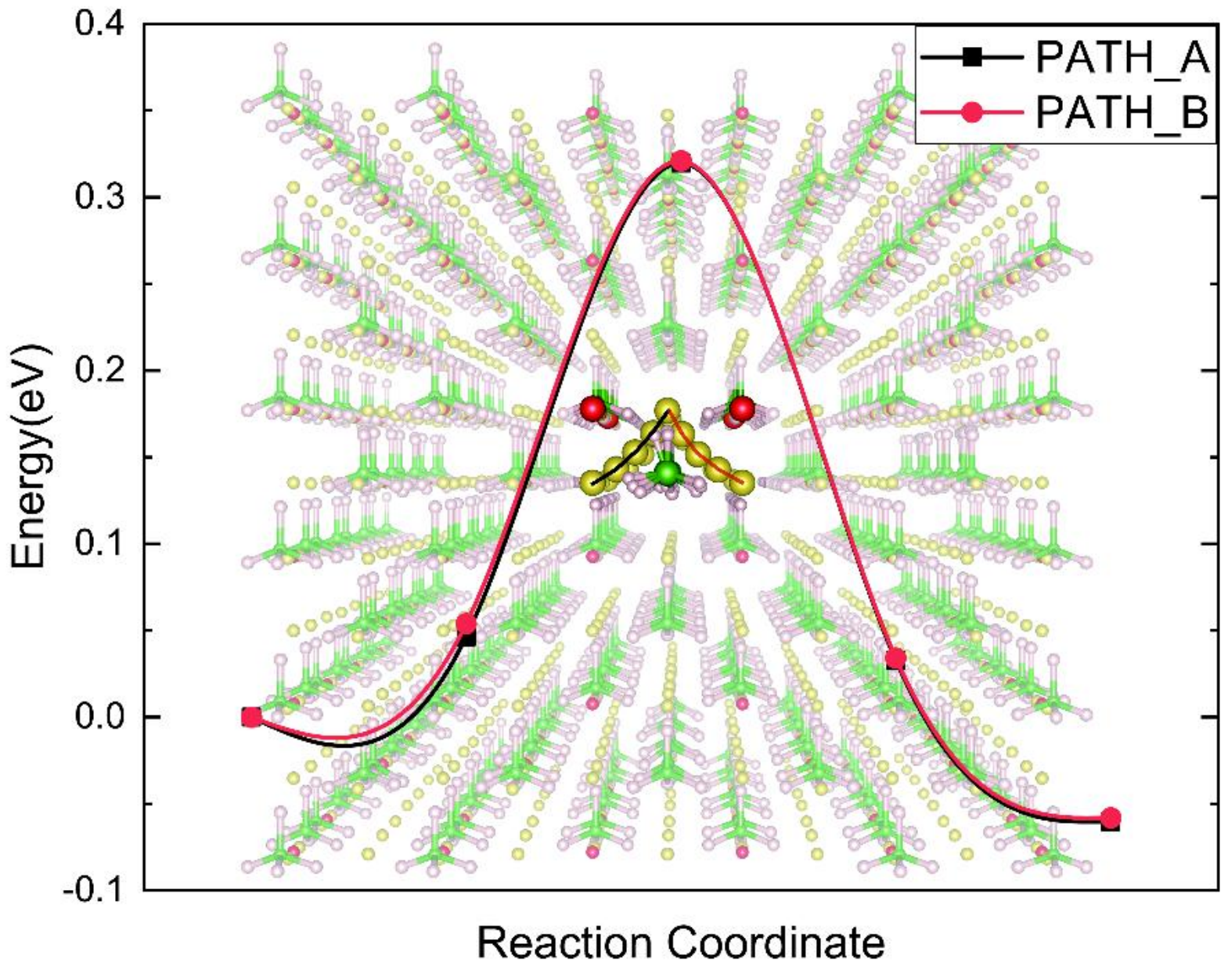

Figure.1 The optimized structures of (a) $Na_4OBr_2$ in *I*4/*mmm*, (b) $Na_4O(BH_4)_2$ in *Pnnm*, (c) $Na_4O(BF_4)_2$ and (d) $Na_4O(AlH_4)_2$ in *Fmm*2 in tetragonal and orthorhombic symmetry.

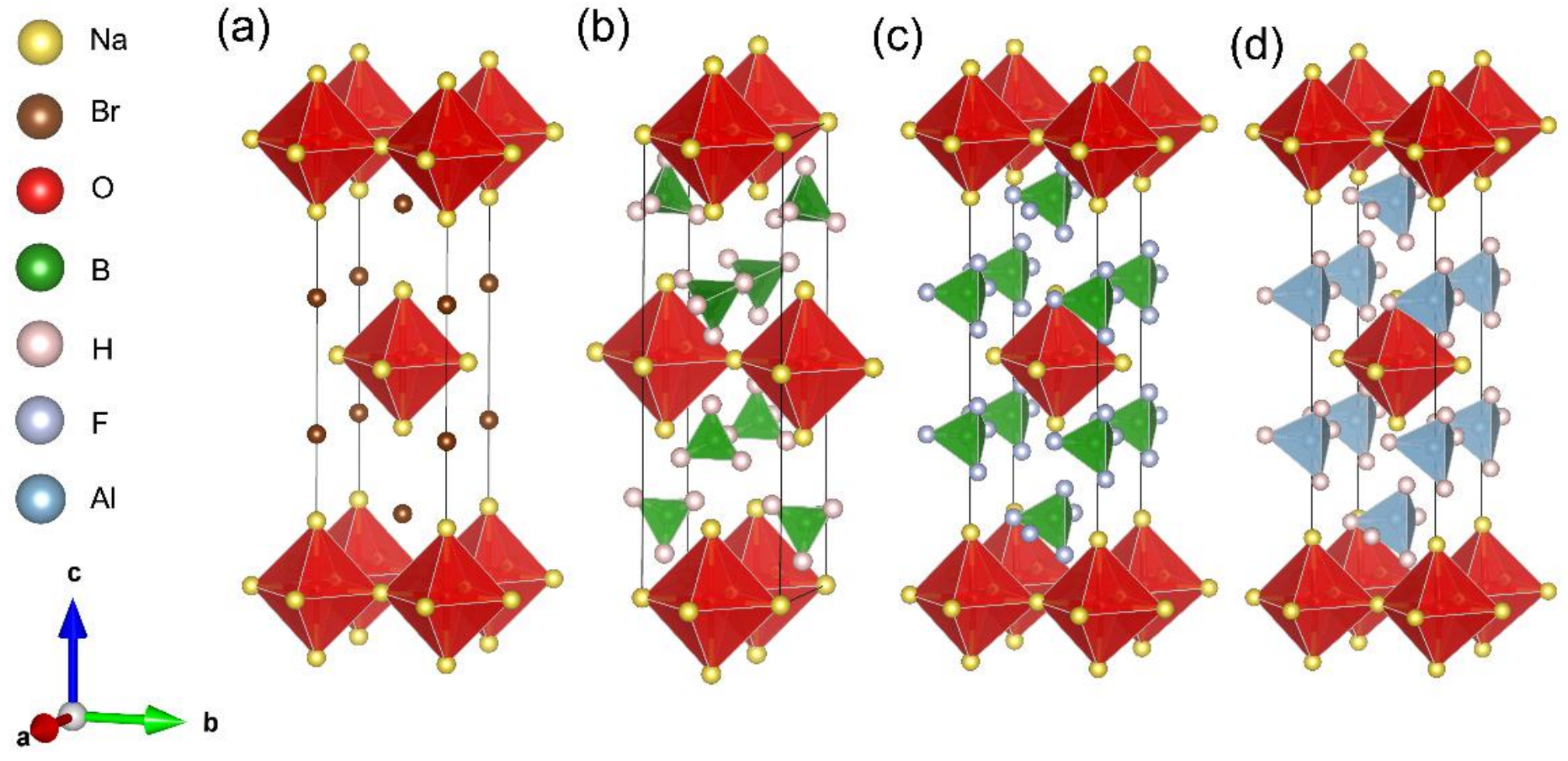

Figure.2 The phonon spectrum and corresponding projection phonon density-of-state of (a)$Na_4OBr_2$, (b)$Na_4O(BH_4)_2$, (c)$Na_4O(BF_4)_2$ and (d)$Na_4O(AlH_4)_2$, respectively.

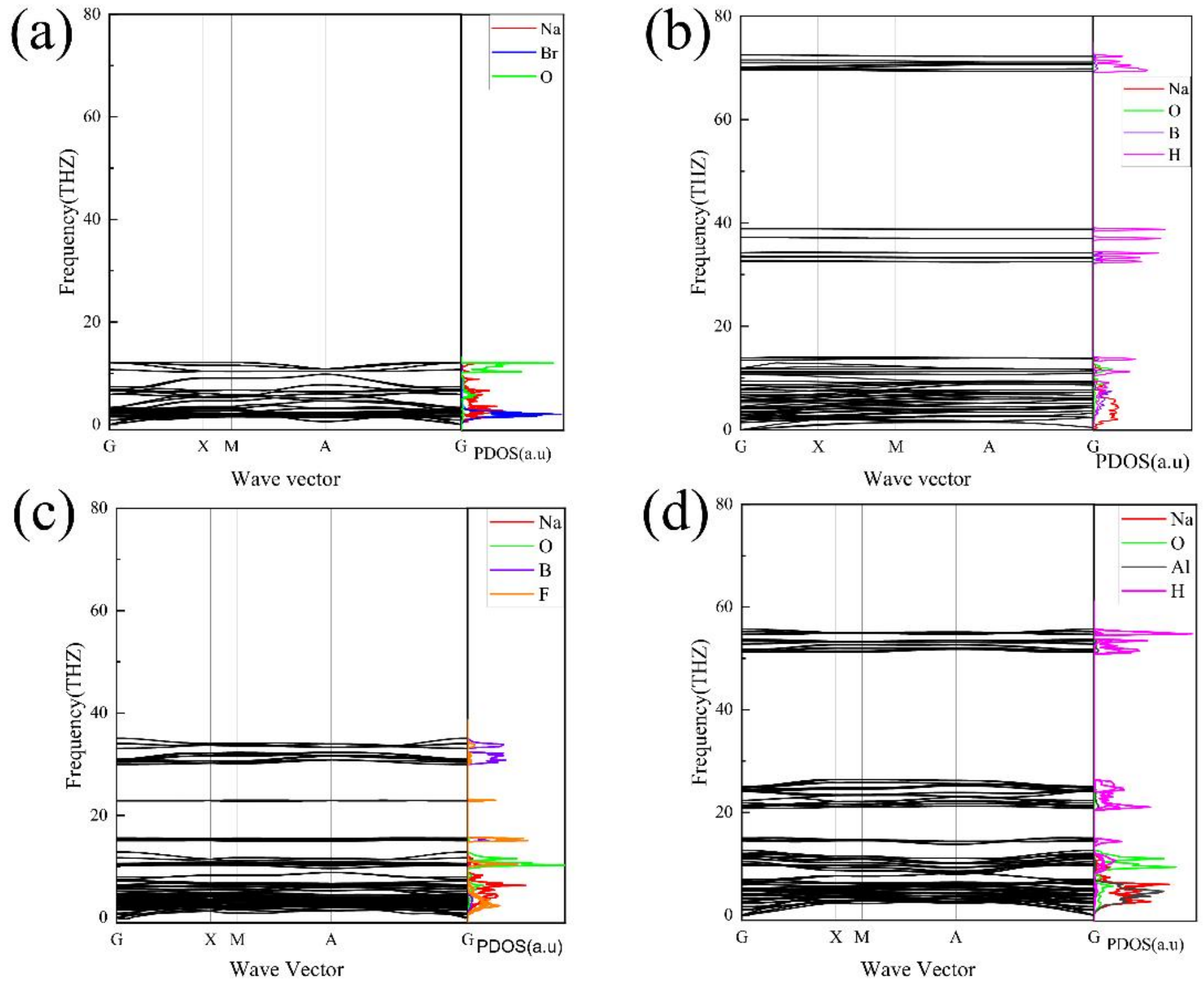

Figure.3 The band structures and the corresponding density of states of (a) $Na_4OBr_2$, (b) $Na_4O(BH_4)_2$, (c) $Na_4O(BF_4)_2$and (d) $Na_4O(AlH_4)_2$, respectively.

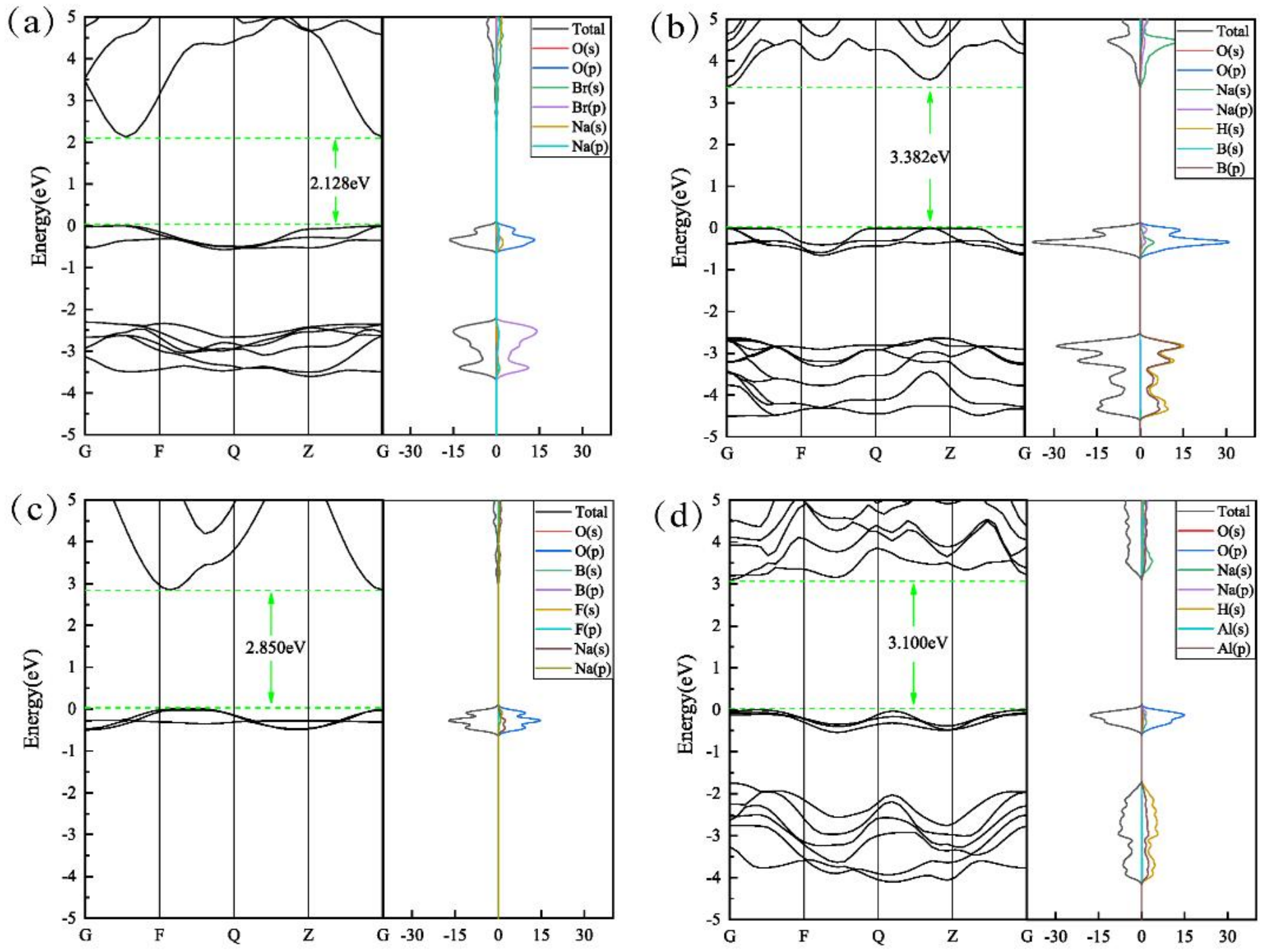

Figure.4 The Crystal Orbital Hamiltonian Populations (COHP) analysis for B-H, B-F, and Al–H bonds in the anion groups.

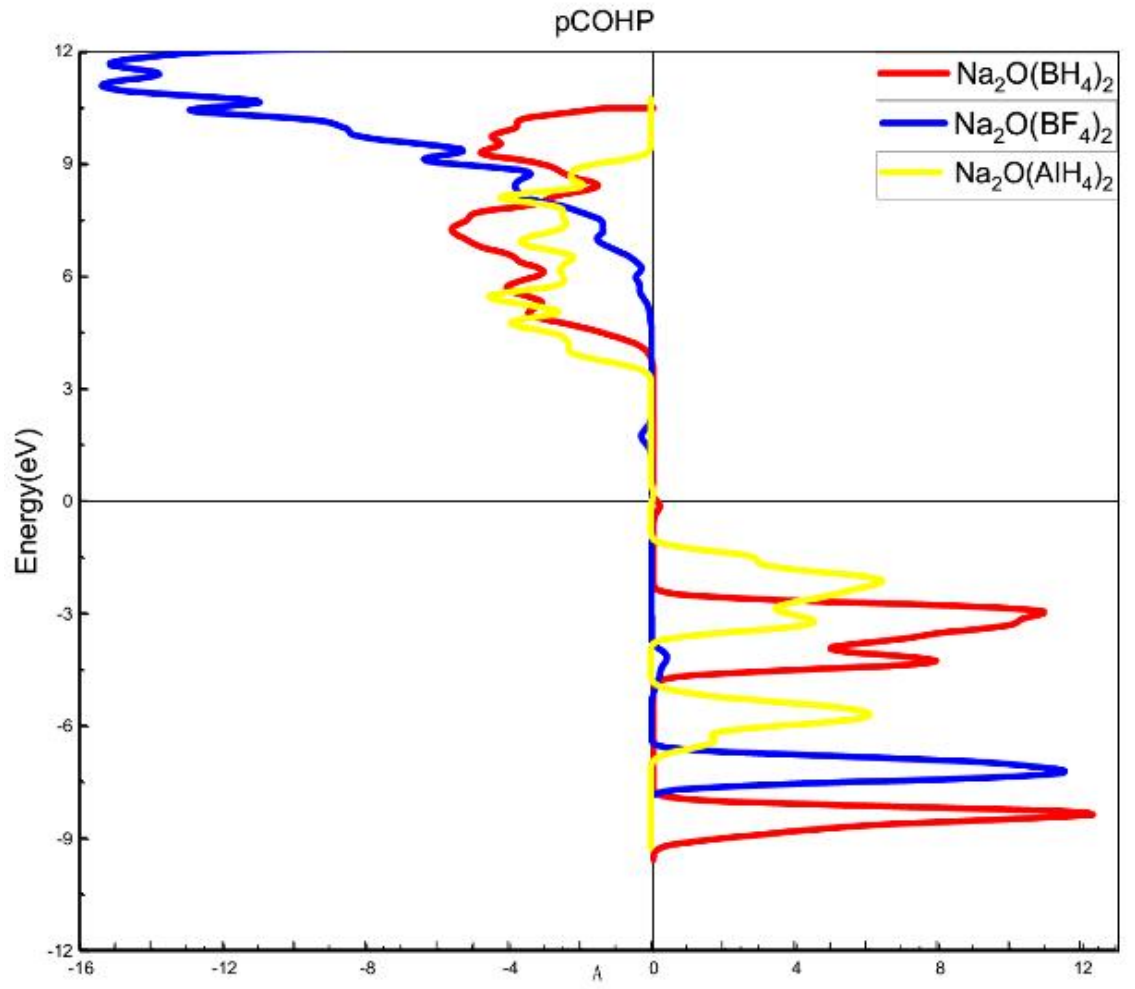

Figure. 5 Taking $Na_4O(BH_4)_2$ as an example, the migration paths of $Na^+$ in paths A and B are shown in red and blue curves in the structure, respectively. The migration mode of other $Na_4OM_2$ is similar. The lines connecting the points correspond to a spline fitted to the calculated CI-NEB energies.

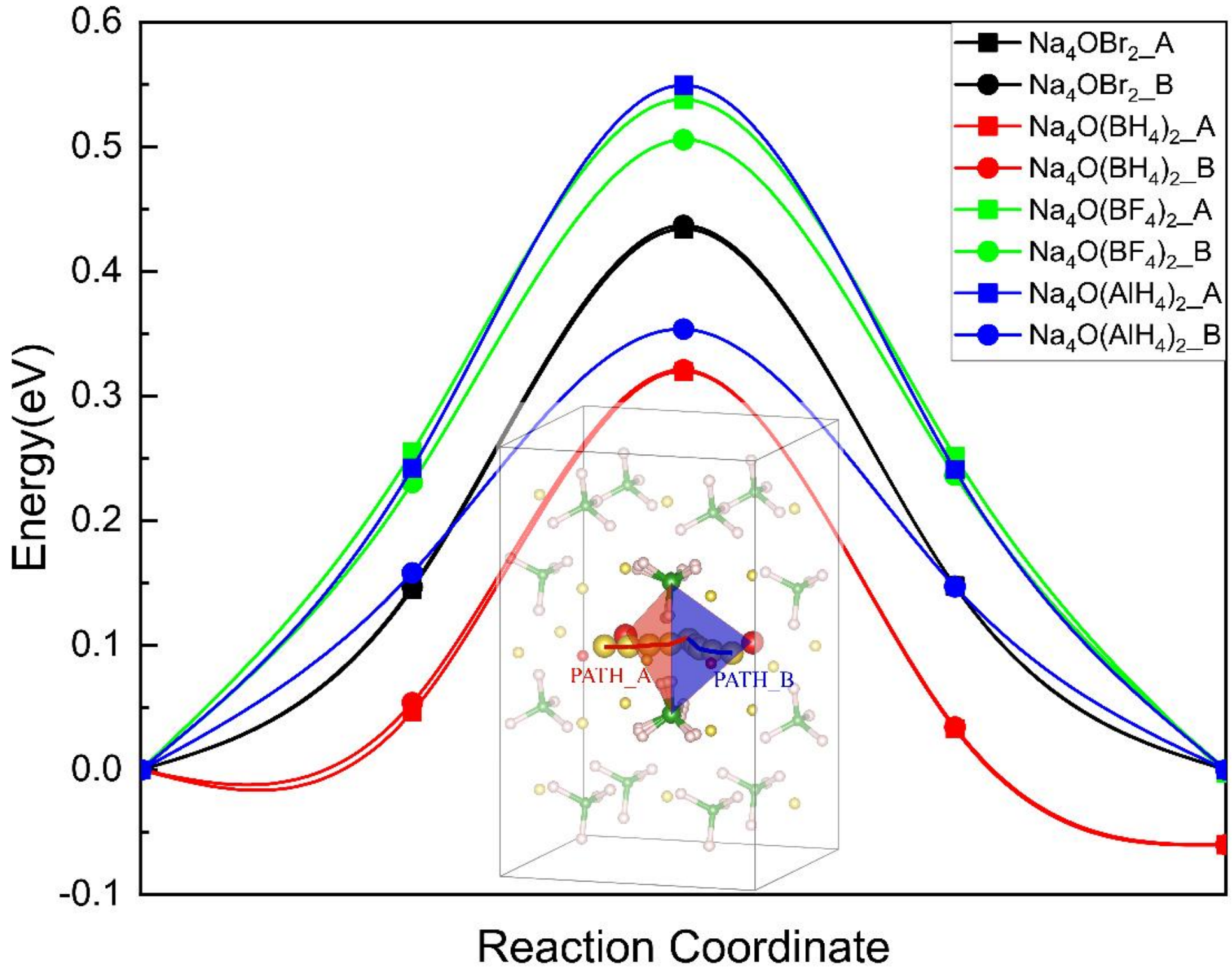

Figure 6. $Na^+$ cations migrate through the MGs and the paddle wheel effect of anion groups in (a) $Na_4OBr_2$, (b) $Na_4O(BH_4)_2$ (c) $Na_4O(BF_4)_2$ and (d) $Na_4O(AlH_4)$ structures.

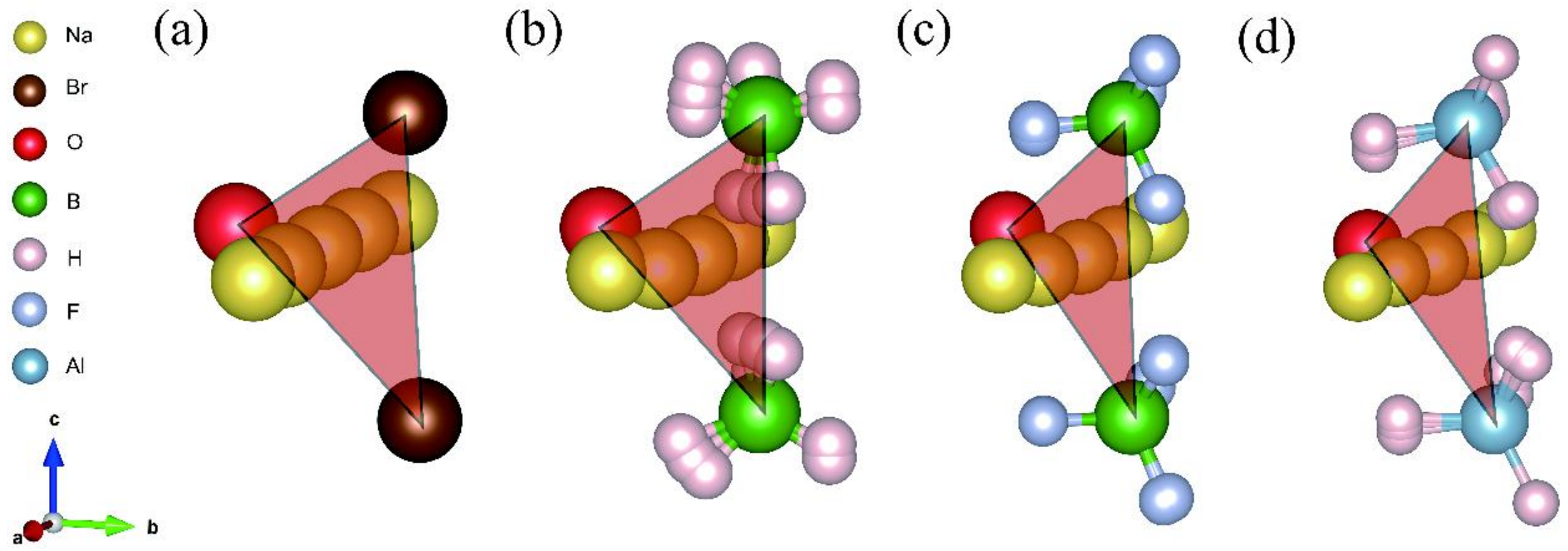

Figure. 7. Schematic diagrams of the coordination of charge carriers in the initial, transition and final states are shown for (a) $Na_4OBr_2$ and (b) $Na_4O(BH_4)_2$ structures. (c) and (d) show the coordination of charge carriers in the initial, transition and final states of $Na_4O(BF_4)_2$ and $Na_4O(AlH_4)$, respectively. Anionic groups with strong interaction and large volume form different initial, transition, and final state energy environments along different pathways. The substituting of $[BF_4]^-$ and $[AlH_4]^-$ results in different energy barriers in the migration path of the $Na_4OM_2$ structure.

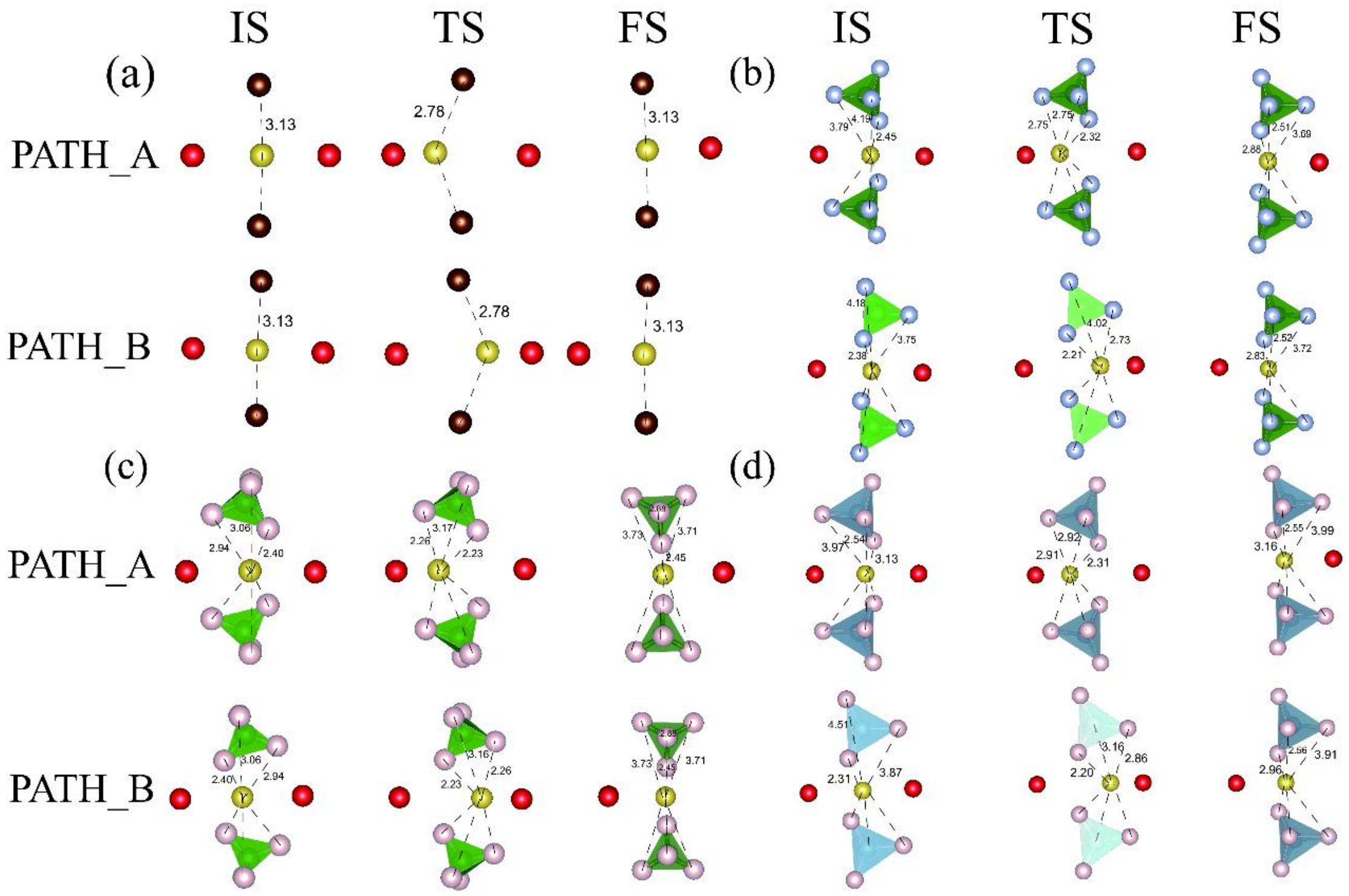

Table 1. Optimized geometries of $Na_4OM_2$ with the PBE-D3.

| | Space group | *a*(Å) | *b*(Å) | *c*(Å) | Atoms pairs | Average distance (Å) |
|---|---|---|---|---|---|---|
| [a] $Na_4OBr_2$ | *I*4/*mmm* (#139) | 4.52 | 4.52 | 14.91 | Na-Br | 3.19 |
| | | | | | Na-O | 2.26 |
| [b] $Na_4OBr_2$ | *I*4/*mmm* (#139) | 4.50 | 4.50 | 14.93 | Na-Br | 3.18 |
| | | | | | Na-O | 2.25 |
| $Na_4O(BH_4)_2$ | *Pnnm* (#58) | 4.46 | 4.46 | 14.67 | Na-B | 2.96 |
| | | | | | B-H | 1.22 |
| | | | | | Na-O | 2.28 |
| $Na_4O(BF_4)_2$ | *Fmm*2 (#42) | 4.64 | 4.57 | 16.23 | Na-B | 3.09 |
| | | | | | B-F | 1.42 |
| | | | | | Na-O | 2.26 |
| $Na_4O(AlH_4)_2$ | *Fmm*2 (#42) | 4.67 | 4.68 | 16.87 | Na-Al | 3.13 |
| | | | | | Al-H | 1.63 |
| | | | | | Na-O | 2.28 |

a: Experimental values from Ref. 25.

b: Optimized lattice parameters.

Table 2. Tolerance factor *t* and octahedral factors μ of $Na_4OM_2$, respectively.

| | Ionic radius of $M^-$ (Å) | Tolerance factor *t* | octahedral factors μ |
|---|---|---|---|
| $Na_4OBr_2$ | 1.96 | 0.87 | 0.71 |
| $Na_4O(BH_4)_2$ | 2.03 | 0.89 | 0.69 |
| $Na_4O(BF_4)_2$ | 2.43 | 1.01 | 0.58 |
| $Na_4O(AlH_4)_2$ | 2.66 | 1.07 | 0.53 |

In each case, the used Shannon effective ionic radii (R) of $Na^+$, $O^{2-}$, B, Al, H, and F are 1.02, 1.40, 0.41, 0.68, 0.81 and 1.01 Å, respectively. [28,42] The radius of the $XY_4^-$ group is defined as the sum of the bond length of X-Y (X = B, Al; Y = H, F) and the ionic radius of Y. The bond length is taken as the optimized average bond length inside the superhalogen, as listed in Table 1.

Table 3. The calculated elastic constants (in Kbar) and Bulk moduli *B*, Shear moduli *G* and Young's moduli *E* (in Gpa) of $Na_4OM_2$.

| | $Na_4OBr_2$ | $Na_4O(BH_4)_2$ | $Na_4O(BF_4)_2$ | $Na_4O(AlH_4)_2$ |
|---|---|---|---|---|
| $C_{11}$ | 509.33 | 636.40 | 715.49 | 392.95 |
| $C_{12}$ | 104.52 | 132.00 | 215.34 | 112.44 |
| $C_{13}$ | 136.58 | 151.86 | 200.91 | 68.98 |
| $C_{22}$ | - | 660.52 | 712.46 | 398.47 |
| $C_{23}$ | - | 146.91 | 197.32 | 72.64 |
| $C_{33}$ | 492.76 | 586.98 | 1014.46 | 487.08 |
| $C_{44}$ | 145.68 | 141.67 | 220.02 | 171.60 |
| $C_{55}$ | - | 172.44 | 200.43 | 100.67 |
| $C_{66}$ | 111.47 | 179.78 | 189.60 | 98.81 |
| $B_{Hill}$ | 25.18 | 30.48 | 77.55 | 19.53 |
| $G_{Hill}$ | 14.51 | 19.13 | 31.80 | 13.63 |
| $E_{Hill}$ | 36.52 | 47.46 | 82.49 | 39.65 |

Table 4. Bader charge analysis and the charge transfer of each ion in $Na_4OM_2$.

| System | Na (*e*) | O (*e*) | Br (*e*) | B (*e*) | Al (*e*) | H (*e*) | F (*e*) | Anion groups (*e*) |
|---|---|---|---|---|---|---|---|---|
| $Na_4OBr_2$ | +0.82 | -1.53 | -0.88 | - | - | - | - | -0.88 |
| $Na_4O(BH_4)_2$ | +0.83 | -1.53 | - | +1.62 | - | -0.63 | - | -0.90 |
| $Na_4O(BF_4)_2$ | +0.83 | -1.54 | - | +2.40 | - | - | -0.83 | -0.92 |
| $Na_4O(AlH_4)_2$ | +0.83 | -1.52 | - | - | +2.14 | -0.76 | - | -0.91 |

Table 5. Ionic conductivity and related parameters of each candidate.

| System | Path | Distance (Å) | Energy Barrier (eV) | Ionic Conductivity ($S\cdot cm^{-1}$) |
|---|---|---|---|---|
| $Na_4OBr_2$ | A | 2.921 | 0.43 | $1.48\times10^{-6}$ |
| | B | 2.920 | 0.43 | $1.34*10^{-6}$ |
| $Na_4O(BH_4)_2$ | A | 2.731 | 0.32 | $1.13*10^{-4}$ |
| | B | 2.727 | 0.32 | $1.09*10^{-4}$ |
| $Na_4O(BF_4)_2$ | A | 3.115 | 0.54 | $8.16*10^{-8}$ |
| | B | 3.007 | 0.50 | $8.88*10^{-8}$ |
| $Na_4O(AlH_4)_2$ | A | 3.359 | 0.55 | $1.93*10^{-8}$ |
| | B | 2.949 | 0.35 | $2.90*10^{-5}$ |

Table 6. The ICOHP of carriers and their neighbors in the transition states during migration

| Atom pair | $d_{Na\text{-}H/F}$_A | $d_{Na\text{-}H/F}$_B | ICOHP_A | ICOHP_B |
|---|---|---|---|---|
| Na-H($Na_4O(BH_4)_2$) | 2.39 | 2.34 | -0.10 | -0.13 |
| Na-F($Na_4O(BF_4)_2$) | 2.51 | 2.38 | -0.27 | -0.38 |
| Na-H($Na_4O(AlH_4)_2$) | 2.54 | 2.31 | -0.17 | -0.09 |